\documentclass[a4paper]{article}
\usepackage{epsfig,amsmath,amsfonts,amssymb,setspace,multirow,textcomp}
\usepackage[T1]{fontenc}
\usepackage{xcolor}
\usepackage{booktabs}
\usepackage{fancyhdr} 
\usepackage{hyperref} 
\usepackage{footnote} 

\makeatletter
\renewcommand{\@oddhead}{\textit{Advances in Astronomy and Space Physics} \hfil}
\renewcommand{\@evenfoot}{\hfil \thepage \hfil}
\renewcommand{\@oddfoot}{\hfil \thepage \hfil}
\makeatother

\renewenvironment{thebibliography}[1]{\begin{oldthebibliography}{#1}\setlength{\parskip}{0ex}\setlength{\itemsep}{0ex}}{\end{oldthebibliography}}

\begin{document}
\fontsize{11}{11}\selectfont 
\title{The search for NEOs as potential candidates for use in space missions to Venus and Mars}
\author{\textsl{A.\,S.~Kasianchuk$^{1}$, V.\,M.~Reshetnyk$^{1}$}}
\date{\vspace*{-6ex}}
\maketitle
\begin{center} {\small $^{1}$Taras Shevchenko National University of Kyiv, Glushkova ave., 4, 03127 Kyiv, Ukraine\\
{\tt kasancukarsen@gmail.com}}
\end{center}

\begin{abstract}
In this work, we analyzed the orbits of more than 35,000 (for 2024) near-Earth objects (NEOs) for the possibility of successive approaches to all pairs of planets: Earth, Venus, and Mars in the time range from 2020 to 2120. We have selected 120 candidates for Earth-Mars, Earth-Venus, Mars-Earth, Mars-Venus, Venus-Earth, and Venus-Mars fast transfers (within 180 days); 2 candidates for double transfers (consecutive approaches with three planets); 10 candidates for multiple transfers, when an asteroid has several consecutive paired approaches to planets in a hundred years.\\[1ex]
\end{abstract}

\section*{\sc introduction}
\indent \indent The effect of space radiation on the health of crew members remains one of the most uncertain issues for long-duration manned space flights. In terms of radiation protection for humans in interplanetary space, the two significant sources of radiation for planetary missions include: heavy ions of the galactic cosmic rays (GCR) and sporadic production of energetic protons from large solar particle events (SPE)\cite{rapp23}. 

Francis Cucinotta et al. \cite{cucin14} note the possible destructive effect of cosmic radiation on the central nervous system (CNS) of a person: 
"Possible CNS risks during a mission are altered cognitive function, including detriments in short-term memory, reduced motor function, and behavioral changes, which may affect performance and human health."

Passive shielding is a promising and currently the only technologically simple solution to the problem of cosmic radiation. Since the budget of the mission will greatly increase if passive shielding made of aluminum or other material is included in the structure of the ship (the weight characteristics of which are limited), it is worth seriously considering an alternative approach -- near-Earth objects (NEOs).

Gregory Matloff et al. \cite{matl11} suggested using the NEOs as passive shielding, in a way where the spacecraft spends most of its flight inside the NEO. They find six candidates for Earth/Mars and Mars/Earth transfers for six hypothetical missions. The idea of finding "reusable" candidates for interplanetary missions is also proposed.

In this work, we analyzed the orbits of more than 35,000 NEOs (for 2024) for the possibility of successive approaches to all pairs of planets: Earth, Venus, Mars. We distinguish three types of transfers: "fast", "double" and "multiple".

\section*{\sc data and methods}

\indent \indent The ephemeris for NEOs were generated using the API of the Jet Propulsion Laboratory (JPL) Horizons online service \footnote{\href{https://ssd.jpl.nasa.gov/horizons/\#api}{https://ssd.jpl.nasa.gov/horizons}}. We generated files for asteroids only, excluding 206 known near-Earth comets due to their uncertain possible activity during the mission. At the output for one asteroid, we had three files in Earth-centric, Venus-centric, and Mars-centric geometric Cartesian coordinate systems.

The files contained NEO's position points (X, Y, Z) for the time range from 2020-Jan-01 to 2120-Jan-01 with step-size in 10 days. In addition to contained, each data step included: ($\mathrm{V_{X}, V_{Y}, V_{Z}}$) -- velocity components (km $\mathrm{s^{-1}}$), LT -- one-way Newtonian light-time (sec), RG -- the distance from the coordinate center (km), RR -- radial velocity with respect to the coordinate center (km $\mathrm{s^{-1}}$) and Julian Day (JD) number.

The method of searching for consecutive close approaches of the asteroid to the planets (the starting planet and the finishing planet) was to find the smallest time difference (transfer time) between the local minima of the asteroid's RG values in the selected planet-centric coordinate systems. The local RG minima were chosen according to a criterion that set the maximum distance to the planet within which the asteroid was considered a potential candidate (d-criterion). We put the d-criterion to the average distance to the Moon. We took one Lunar distance (LD) = 385,000 km.

We were looking for candidates for three types of transfers:
\begin{enumerate}
    \item "fast":

    d-criterion = 25 LD

    max. transfer time = 180 days
    
    \item "double":

    d-criterion = 25 LD

    max. transfer time = 365 days
    
    \item "multiple":

     d-criterion = 40 LD

     max. transfer time = 250 days 
\end{enumerate}

Despite the fact that the temporal criteria we have chosen do not serve any specific physical goals other than to characterize the three types of transfers we have proposed, the d-criteria were chosen for the relevant reasons: If we expect that Potentially Hazardous Asteroids (PHA) will have the highest probability of being selected, NEAs whose Minimum Orbit Intersection Distance (MOID) with the Earth is 0.05 AU or less \footnote{\href{https://cneos.jpl.nasa.gov/about/neo_groups.html}{https://cneos.jpl.nasa.gov/about/neogroups.html}}, and convert 0.05 AU to lunar distances, we get about 19.4 LD. Since most of the asteroids in the PHA list do not have an uncertainty parameter "U" 0, and to ensure that we cover all PHAs with an additional reserve, we added 5.6 LD to the d-criterion. As for the d-criterion of 40 LD, it was chosen based on concerns that we would not be able to find enough asteroids for clarity for "multiple" transfers. That's why we decided to take about twice the distance for 0.05 AU.

\begin{figure}[!h]
\centering
\begin{minipage}[t]{1\linewidth}
\centering
\epsfig{file = 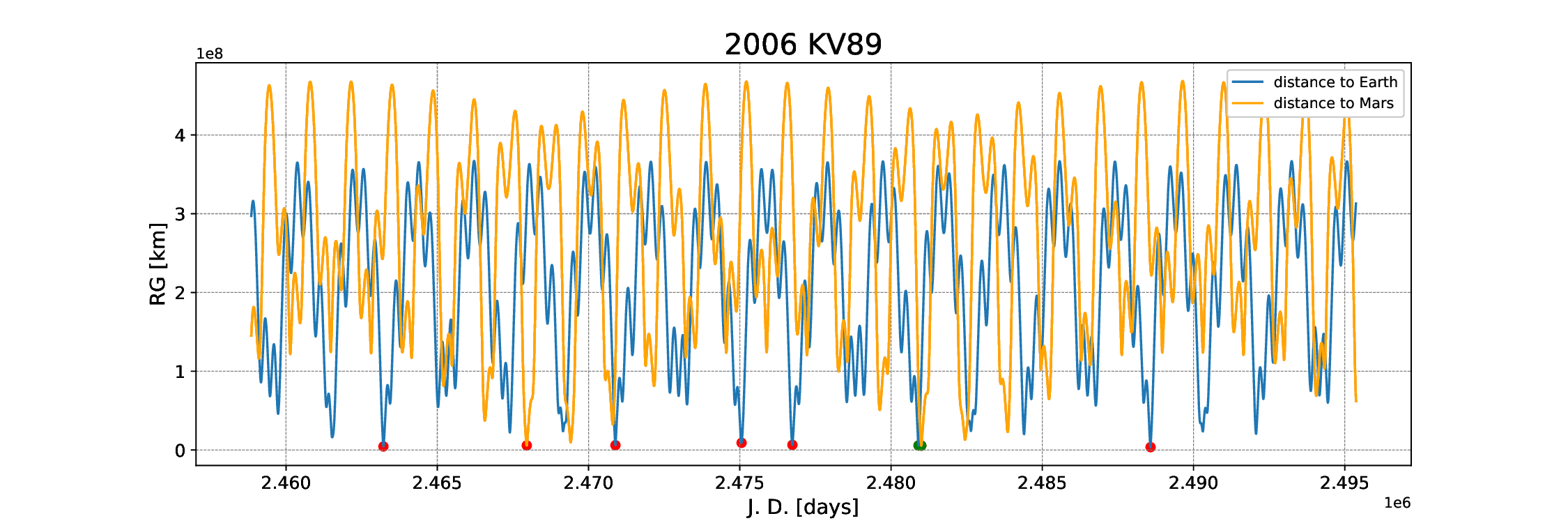,width = 1\linewidth}
\caption{RG curves to Earth \& to Mars for  2006 KV89.}\label{fig1}
\end{minipage}
\end{figure}

Fig.~\ref{fig1} shows the result of finding the transfer time for the asteroid 2006 KV89, which was selected as a candidate for a possible mission from Earth to Mars (start at 2080-May-25 \& finish at 2080-Aug-23). This figure also describes the search method. 

The red points are the local RG minima that passed the d-criterion. Green points are selected minima with the smallest time difference (transfer time) and describe the start and finish of the transfer.

After obtaining the results for a particular way (e.g. Earth-Mars), we conducted an additional selection by the uncertainty parameter "U" \footnote{\href{https://www.minorplanetcenter.net/iau/info/UValue.html}{https://www.minorplanetcenter.net/iau/info/UValue.html}}, where we kept only candidates with values from 0 to 5. 

As shown in the Tsiolkovsky’s rocket equation (Equation~\ref{form1} \cite{ver18}), a linear change in delta-V produces an exponential change in rocket propellant. Therefore, we also made an additional selection based on the asteroid's speed near the planets, which should not exceed 30 km $\mathrm{s^{-1}}$. This was added to weed out missions that were too "expensive".
\begin{equation}\label{form1}
\frac{M_i}{M_f} = \exp{\frac{\Delta V}{w}}
\end{equation}
Where $M_i$ is the initial mass of the rocket, $M_f$ is the final mass of the rocket when the rocket has reached the mission velocity $\Delta V$, and $w$ is the exhaust velocity.

\section*{\sc results and discussion}

\indent \indent We got 525 candidates for "fast" missions without additional selection. The selection by speed during approach and the "U" parameter reduced the list to 120 candidates.

\begin{figure}[!h]
\centering
\includegraphics[width=1\linewidth]{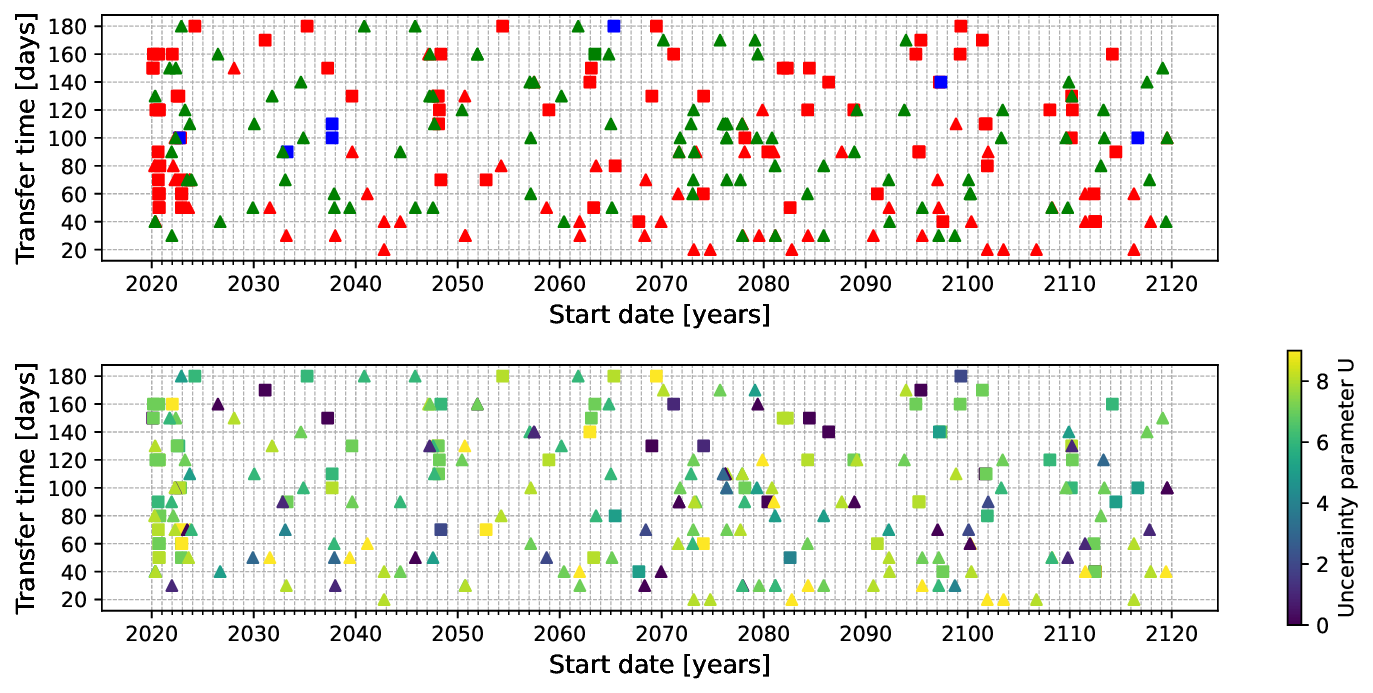}
\caption{NEOs for missions from Earth. In both charts, the square mark corresponds to the Earth-Mars transfer, and the triangular Earth-Venus. In the upper chart, the colour indicates the NEO class: Apollos (red), Amors (blue), Atens (green), Atiras (violet).}\label{fig2}
\end{figure}

For Earth-Venus 162 candidates, 44 candidates after selection; for Earth-Mars 87 candidates, 17 candidates after selection. Fig.~\ref{fig2} shows all possible transfers that will start near the Earth within a radius of 25 LD, counting from the Earth's center. The candidates after additional selection are listed in Tables~\ref{tab1}-\ref{tab2} in Appendix 1. 

\begin{figure}[!h]
\centering
\includegraphics[width=1\linewidth]{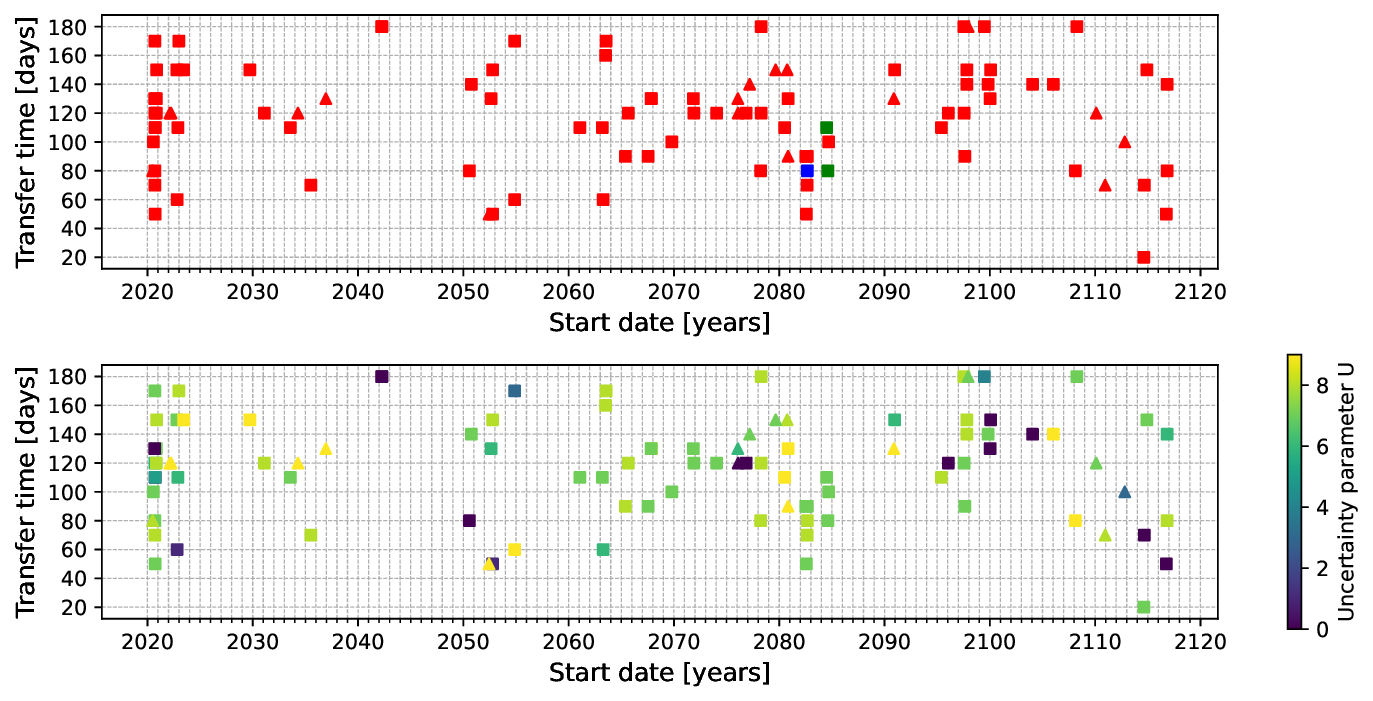}
\caption{NEOs for missions from Mars. In both charts, the square mark corresponds to the Mars-Earth transfer, and the triangular Mars-Venus. In the upper chart, the colour indicates the NEO class: Apollos (red), Amors (blue), Atens (green), Atiras (violet).}\label{fig3}
\end{figure}

For Mars-Earth 79 candidates (17 with "U" 0-5), 13 candidates after selection; for Mars-Venus 17 candidates, 2 candidates after selection. Fig.~\ref{fig3} shows all possible transfers that will start near the Earth within a radius of 25 LD, counting from the Earth's center. The candidates after additional selection are listed in Tables~\ref{tab3}-\ref{tab4} in Appendix 1.

\begin{figure}[!h]
\centering
\includegraphics[width=1\linewidth]{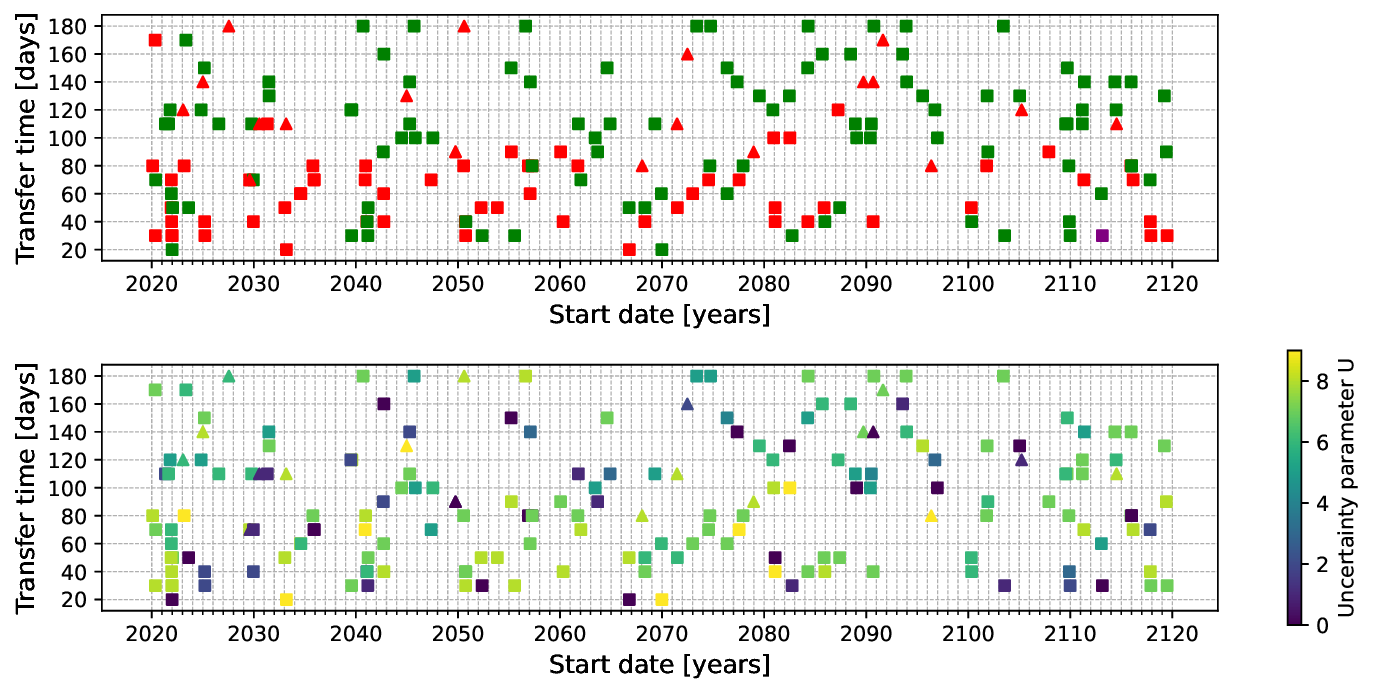}
\caption{NEOs for missions from Venus. In both charts, the square mark corresponds to the Venus-Earth transfer, and the triangular Venus-Mars. In the upper chart, the colour indicates the NEO class: Apollos (red), Amors (blue), Atens (green), Atiras (violet)}\label{fig4}
\end{figure}

For Venus-Earth 161 candidates, 38 candidates after selection; for Venus-Mars 19 candidates, 6 candidates after selection. Fig.~\ref{fig4} shows all possible transfers that will start near the Earth within a radius of 25 LD, counting from the Earth's center. The candidates after additional selection are listed in Tables~\ref{tab5}-\ref{tab6} in Appendix 1.

More and more NEOs are found every year. Therefore, the number of candidates for interplanetary missions is not limited to our tables and will grow every year.

Many asteroids that could have been included in the main candidate tables were not selected due to the high uncertainty parameter of the orbit "U". Therefore, in Tables~\ref{tab7}-\ref{tab12} we present limited lists of asteroids with "U" 6-7 that are possible candidates. Table~\ref{tab5} does not contain goals with "U" 5, but they are listed in Table~\ref{tab11}. These asteroids require additional astrometric measurements of the orbit, and hence observations. Some of them are good candidates for missions, perhaps even better than the previous ones, and some will not be included in the new lists. No asteroid with "U" 6-7 was found for Venus-Mars transfer.

We pay special attention to the promising idea of "multiple" transfers. We have found 11 candidates for missions between the planets Earth and Venus, they are listed in Tables~\ref{tab15}-\ref{tab16}. Also, there was only one candidate for the Venus-Mars transfer, see Table~\ref{tab17}. Unfortunately, there were no asteroids with more than three flights that met the selection criteria for the other possible flight options during the period of time we studied.

\newpage

\newpage
In the following, we estimate the diameter of some asteroids from the multiple transfer lists. We simply use an assumed mean albedo of 14\% and absolute magnitude H data from the JPL databases \footnote{\href{https://ssd.jpl.nasa.gov/tools/sbdb_lookup.html\#/}{https://ssd.jpl.nasa.gov/tools1}}, as does the Center for NEO Studies (CNEOS) \footnote{\href{https://cneos.jpl.nasa.gov/stats/}{https://cneos.jpl.nasa.gov/stats}}.
The expression for diameter $D$ in km as $a$ function of absolute magnitude $H$ and geometric albedo a is given by the following equation \cite{bow89} \cite{harr97}.

\begin{equation}\label{form2}
D = 10^{[3.1236 - 0.5\log_{10}(a) - 0.2H]}
\end{equation}

The Earth-Venus and Venus-Earth lists include interesting objects. The largest estimated size is the 2014 QX432 with a diameter of 0.370 km. Next are 2009 WY7 (0.054 km), 2018 RY1 (0.046 km), 2011 EX4 (0.043 km), 2011 EP51 (0.032 km), and 2023 JK3 (0.030 km). Asteroid 2019 SF6, which could be used as a transfer for 13 missions, has a diameter of 0.020 km. Next are 2005 VL1 (0.018 km), 2023 TM3 (0.015 km). The candidate for the Venus to Mars transfers 2019 YU3 has a diameter of 0.100 km. All these estimates are approximate and subject to a significant margin of error.

The reader can estimate the diameter of the asteroid of interest using this CNEOS website. \footnote{\href{https://cneos.jpl.nasa.gov/tools/ast_size_est.html}{https://cneos.jpl.nasa.gov/tools2}}

We found only 2 candidates for "double" transfers Earth-Venus-Mars and Mars-Venus-Earth. Information about them can be found in the Table~\ref{tab13}-\ref{tab14} in Appendix 2. Asteroids 2016 DY30 and 2005 TK50 could be good for such missions, but they have too high a "U" (7 \& 9) and a small diameter (0.003 km and 0.005 km) relative to the expected size of a spacecraft with a crew. However, it is worth mentioning them to emphasise the rarity of such transfers last less than a year.

\section*{\sc conclusions}
\indent \indent This study has shown that the use of NEOs as shielding during space flights is promising and can be considered when planning to send the first astronauts to Mars. Given that a human mission to Mars is a highly complex problem \cite{rapp23} (this also applies to Venus), and that establishing a radiation protection station in NEO may encounter several technological problems that are difficult to predict now, it is worth choosing an initial NEO target from the list for one-time "fast" missions.

At present, we have a relatively extensive list of targets for multiple transfers (which are expected to become the main ones for future flights) only for the Earth-Venus and Venus-Earth routes. Over time, the discovery of new NEOs, as well as the reduction of the "U" uncertainty of the orbit of the known targets listed in this article (see the lists in Tables~\ref{tab7}-\ref{tab12}), will lead to the replenishment of the lists of multiple interplanetary transfers.  It also puts even more hope in NASA's NEO Surveyor mission, which has the goal of "finding more than 90 per cent of all NEOs larger than 140 metres in diameter" \footnote{\href{https://www.jpl.nasa.gov/missions/near-earth-object-surveyor}{https://www.jpl.nasa.gov/missions}}.

Asteroids such as 2014 QX432, 2009 WY7, and 2019 YU3 as well as estimates of their size, clearly prove the benefits of their use. At the same time, when choosing a NEO as a radiation shield, one should not only focus on its size, but also on the minimum distance of its approach to the target planets. In this article, we set the d-criterion at 40 LD to demonstrate the prospects of using NEOs for multiple shielding. If a spacecraft with a crew of astronauts stays in open space, outside the dominance of the Earth's magnetic field, for a longer period of time than inside the NEO, the question arises as to the effectiveness and meaning of the method. Therefore, the d-criterion should be narrowed in the future.


\newpage
\section*{\sc appendix 1}

\begin{table}[!h]
 \centering
 \caption{Candidates for missions from Earth to Venus. Where "NEO" - asteroid name, "U" - uncertainty parameter of the orbit, "Tr. time" - transfer time, "Start time/Finish time" - date of approach to the launch/final planet, "$\mathrm{r_{1/2}}$" - distance to the launch/final planet during the approach, "$\mathrm{V_{1/2}}$" - target speed at the point of launch/final approach.}\label{tab1}
 \vspace*{2ex}
\begin{tabular}{@{}ccccccccc@{}}
\toprule
NEO      & U & Tr. time (days) & Start time  & $\mathrm{r_{1}}$ (km)     & $\mathrm{V_{1}}$ (km $\mathrm{s^{-1}}$) & Finish time & $\mathrm{r_{2}}$ (km) & $\mathrm{V_{2}}$ (km $\mathrm{s^{-1}}$) \\ \midrule
1997 NC1   & 0 & 160 & 2026-Jun-28 & 2598054 & 8.89  & 2026-Dec-05 & 4723428 & 9.88  \\
2012 WH1   & 3 & 50  & 2029-Nov-29 & 4439561 & 9.23  & 2030-Jan-18 & 6288297 & 13.39 \\
2019 AR8   & 5 & 110 & 2030-Jan-18 & 5912283 & 7.92  & 2030-May-08 & 4017152 & 8.34  \\
2009 WY7   & 1 & 90  & 2032-Nov-03 & 7482158 & 14.67 & 2033-Feb-01 & 3531904 & 20.79 \\
2005 VL1   & 4 & 70  & 2033-Feb-01 & 3338198 & 5.94  & 2033-Apr-12 & 5108226 & 6.74  \\
2020 UE4   & 3 & 50  & 2037-Nov-27 & 8023461 & 9.62  & 2038-Jan-16 & 6006599 & 12.68 \\
2016 LD9   & 1 & 30  & 2037-Dec-27 & 5482923 & 16.20 & 2038-Jan-26 & 3549524 & 11.61 \\
2019 SF6   & 0 & 50  & 2045-Nov-05 & 8356719 & 8.45  & 2045-Dec-25 & 9111880 & 14.46 \\
2011 GE3   & 1 & 130 & 2047-Mar-30 & 3148473 & 6.96  & 2047-Aug-07 & 7469566 & 8.90  \\
2017 HU2   & 5 & 160 & 2047-Apr-09 & 4921729 & 4.27  & 2047-Sep-16 & 3193152 & 3.84  \\
2015 OQ21  & 3 & 120 & 2050-Jun-02 & 4957528 & 7.80  & 2050-Sep-30 & 5852245 & 11.45 \\
2000 AC6   & 0 & 160 & 2051-Dec-04 & 9415242 & 6.80  & 2052-May-12 & 4907489 & 9.17  \\
2002 XY38  & 0 & 100 & 2057-Feb-25 & 7087027 & 8.02  & 2057-Jun-05 & 7189113 & 4.96  \\
2002 LT24  & 0 & 140 & 2057-Jun-25 & 9232247 & 8.95  & 2057-Nov-12 & 7717317 & 18.66 \\
2021 PQ6   & 2 & 50  & 2058-Sep-18 & 7054460 & 10.19 & 2058-Nov-07 & 9125692 & 10.70 \\
2016 WJ1   & 0 & 40  & 2061-Dec-11 & 4281840 & 14.81 & 2062-Jan-20 & 3451518 & 11.64 \\
2009 VW    & 0 & 30  & 2068-Apr-28 & 7500997 & 17.02 & 2068-May-28 & 9129845 & 17.13 \\
2015 KP18  & 2 & 70  & 2068-Jun-07 & 2256323 & 17.37 & 2068-Aug-16 & 8330325 & 11.82 \\
2019 VH5   & 0 & 40  & 2069-Dec-09 & 5773454 & 11.85 & 2070-Jan-18 & 5502383 & 8.67  \\
1999 GS6   & 0 & 90  & 2071-Sep-10 & 3481846 & 14.54 & 2071-Dec-09 & 9612736 & 15.47 \\
2016 BC14  & 1 & 110 & 2076-Mar-27 & 5623184 & 6.50  & 2076-Jul-15 & 4310520 & 6.47  \\
2016 BC14  & 1 & 110 & 2076-Mar-27 & 5623184 & 6.50  & 2076-Jul-15 & 4310520 & 6.47  \\
2016 HD3   & 3 & 100 & 2076-May-06 & 8202987 & 8.64  & 2076-Aug-14 & 5447434 & 5.07  \\
2019 JH7   & 5 & 100 & 2076-May-16 & 654143  & 9.19  & 2076-Aug-24 & 6723272 & 8.83  \\
2014 QX432 & 0 & 30  & 2077-Dec-17 & 8406818 & 18.08 & 2078-Jan-16 & 9491640 & 15.86 \\
2022 DJ1   & 5 & 170 & 2079-Feb-20 & 5375733 & 5.24  & 2079-Aug-09 & 7046225 & 5.68  \\
2023 JK3   & 5 & 100 & 2079-May-01 & 3390197 & 4.47  & 2079-Aug-09 & 3026153 & 8.66  \\
2003 LN6   & 0 & 160 & 2079-May-31 & 3482382 & 3.47  & 2079-Nov-07 & 5855530 & 6.89  \\
2023 TX6   & 5 & 80  & 2081-Feb-09 & 4606209 & 7.81  & 2081-Apr-30 & 6128165 & 5.77  \\
2005 VN5   & 5 & 80  & 2085-Nov-15 & 3860465 & 5.87  & 2086-Feb-03 & 4212704 & 3.43  \\
2015 XF261 & 1 & 50  & 2092-Apr-12 & 7569718 & 8.57  & 2092-Jun-01 & 3189431 & 8.73  \\
2012 QC8   & 0 & 30  & 2098-Sep-28 & 5545913 & 17.73 & 2098-Oct-28 & 9118277 & 12.05 \\
2016 VG    & 4 & 110 & 2098-Nov-07 & 3447850 & 11.22 & 2099-Feb-25 & 5295270 & 13.01 \\
2019 BE5   & 2 & 70  & 2100-Jan-31 & 4305373 & 13.08 & 2100-Apr-11 & 9211407 & 24.76 \\
2011 EP51  & 0 & 60  & 2100-Mar-22 & 6185111 & 7.11  & 2100-May-21 & 5606121 & 11.66 \\
2014 WO371 & 2 & 90  & 2102-Jan-01 & 9612080 & 9.95  & 2102-Apr-01 & 7195147 & 12.72 \\
2020 BS12  & 0 & 20  & 2103-Jul-05 & 8882151 & 25.59 & 2103-Jul-25 & 5812243 & 28.13 \\
2008 BX2   & 1 & 50  & 2108-Mar-30 & 8577431 & 8.30  & 2108-May-19 & 8388067 & 11.23 \\
2021 SU1   & 5 & 140 & 2109-Nov-30 & 7483871 & 4.94  & 2110-Apr-19 & 2118143 & 8.38  \\
2007 EG    & 1 & 130 & 2110-Mar-20 & 3361474 & 8.49  & 2110-Jul-28 & 8487126 & 15.53 \\
2016 NO56  & 1 & 60  & 2111-Jul-03 & 6720961 & 12.86 & 2111-Sep-01 & 9365062 & 5.49  \\
2018 BH3   & 3 & 120 & 2113-Apr-23 & 7393350 & 10.44 & 2113-Aug-21 & 8332342 & 14.49 \\
2000 UK11  & 1 & 70  & 2117-Oct-29 & 2002900 & 6.18  & 2118-Jan-07 & 6321513 & 6.97  \\
2008 NP3   & 0 & 100 & 2119-Jul-21 & 6093681 & 8.77  & 2119-Oct-29 & 6287244 & 10.60 \\  \bottomrule 
\end{tabular}
\end{table}

\begin{table}[!h]
 \centering
 \caption{Candidates for missions from Earth to Mars.}\label{tab2}
 \vspace*{2ex}
\begin{tabular}{@{}ccccccccc@{}}
\toprule
NEO      & U & Tr. time (days) & Start time  & $\mathrm{r_{1}}$ (km)     & $\mathrm{V_{1}}$ (km $\mathrm{s^{-1}}$) & Finish time & $\mathrm{r_{2}}$ (km) & $\mathrm{V_{2}}$ (km $\mathrm{s^{-1}}$) \\ \midrule
2015 TJ1  & 0 & 170                  & 2031-Feb-12 & 4129563 & 3.74                      & 2031-Aug-01 & 2752232 & 5.94                      \\
2007 FF1  & 0 & 150                  & 2037-Apr-01 & 8044773 & 12.69                     & 2037-Aug-29 & 4838836 & 12.38                     \\
2001 BF10 & 2 & 70                   & 2048-May-03 & 2982890 & 9.84                      & 2048-Jul-12 & 3151699 & 13.07                     \\
2024 EJ4 & 5 & 80                   & 2065-Jun-03 & 8133144 & 7.94                      & 2065-Aug-22 & 9199834 & 10.75                     \\
2019 GN4  & 5 & 40                   & 2067-Oct-01 & 6083210 & 19.12                     & 2067-Nov-10 & 7196682 & 14.51                     \\
2000 OM   & 0 & 130                  & 2069-Jan-13 & 8024423 & 19.69                     & 2069-May-23 & 5719557 & 20.01                     \\
2011 CD22 & 1 & 160                  & 2071-Mar-04 & 6457595 & 17.62                     & 2071-Aug-11 & 9457770 & 11.34                     \\
2007 DA   & 1 & 130                  & 2074-Feb-06 & 9298809 & 16.84                     & 2074-Jun-16 & 9210000 & 7.38                      \\
2006 KV89 & 0 & 90                   & 2080-May-25 & 5747035 & 8.69                      & 2080-Aug-23 & 5588168 & 7.04                      \\
2018 EQ1  & 4 & 50                   & 2082-Aug-03 & 9122204 & 11.73                     & 2082-Sep-22 & 9376785 & 13.19                     \\
2011 WU95 & 0 & 150                  & 2084-Jun-23 & 7589061 & 8.78                      & 2084-Nov-20 & 8466454 & 8.53                      \\
2017 TN6  & 0 & 140                  & 2086-May-14 & 9154626 & 17.83                     & 2086-Oct-01 & 7749273 & 15.21                     \\
1998 VD32 & 0 & 170                  & 2095-May-27 & 6550077 & 8.22                      & 2095-Nov-13 & 8801449 & 6.23                      \\
2003 DW10 & 5 & 140                  & 2097-Mar-17 & 7047157 & 6.71                      & 2097-Aug-04 & 7839381 & 9.82                      \\
2017 HB1  & 2 & 180                  & 2099-Apr-26 & 2425751 & 7.38                      & 2099-Oct-23 & 4105230 & 6.25                      \\
2013 TD   & 0 & 110                  & 2101-Sep-13 & 3365380 & 7.43                      & 2102-Jan-01 & 9359908 & 11.99                     \\
2020 BP   & 0 & 40                   & 2112-Jun-17 & 8443241 & 16.11                     & 2112-Jul-27 & 9589970 & 18.45                     \\
2015 RO82   & 5 & 90                   & 2114-Jul-07 & 8784644 & 7.63                     & 2114-Oct-05 & 7144762 & 10.02                     \\
2023 OS8  & 5 & 100                  & 2116-Sep-04 & 6742955 & 7.44                      & 2116-Dec-13 & 4543610 & 11.18                     \\  \bottomrule
\end{tabular}
\end{table}

\begin{table}[!h]
 \centering
 \caption{Candidates for missions from Mars to Earth.}\label{tab3}
 \vspace*{2ex}
\begin{tabular}{@{}ccccccccc@{}}
\toprule
NEO      & U & Tr. time (days) & Start time  & $\mathrm{r_{1}}$ (km)     & $\mathrm{V_{1}}$ (km $\mathrm{s^{-1}}$) & Finish time & $\mathrm{r_{2}}$ (km) & $\mathrm{V_{2}}$ (km $\mathrm{s^{-1}}$) \\ \midrule
2000 GE2   & 0 & 180 & 2042-Apr-05 & 8831501 & 12.05 & 2042-Oct-02 & 4673008 & 15.33 \\ 
2012 TC4   & 0 & 80  & 2050-Aug-01 & 9441647 & 10.42 & 2050-Oct-20 & 1842253 & 6.17  \\
2013 GS66  & 1 & 50  & 2052-Oct-09 & 5976894 & 16.46 & 2052-Nov-28 & 5603066 & 12.13 \\
2013 BV15  & 3 & 170 & 2054-Nov-18 & 9081427 & 10.91 & 2055-May-07 & 5223938 & 7.06  \\
2002 GM2   & 0 & 120 & 2076-Nov-12 & 9339041 & 21.65 & 2077-Mar-12 & 7542435 & 25.27 \\
2015 KO57  & 0 & 120 & 2096-Jan-22 & 8743388 & 24.93 & 2096-May-21 & 8487455 & 23.58 \\
2008 SE85  & 0 & 150 & 2097-Oct-23 & 9414263 & 19.85 & 2098-Mar-22 & 8916714 & 20.22 \\
2022 SO113 & 4 & 180 & 2099-Jun-25 & 8794206 & 3.89  & 2099-Dec-22 & 5694965 & 3.40  \\
2005 WY55  & 0 & 130 & 2100-Jan-11 & 7583941 & 20.70 & 2100-May-21 & 4814202 & 16.84 \\
2002 XR14  & 0 & 150 & 2100-Jan-31 & 8215261 & 19.11 & 2100-Jun-30 & 4165296 & 17.49 \\
2021 FD1   & 0 & 140 & 2104-Jan-21 & 9599187 & 14.93 & 2104-Jun-09 & 9614472 & 10.86 \\
2017 QL33  & 0 & 70  & 2114-Sep-05 & 7955139 & 11.76 & 2114-Nov-14 & 4281736 & 7.31  \\
1998 MZ    & 0 & 50  & 2116-Oct-04 & 9502518 & 15.69 & 2116-Nov-23 & 6035717 & 17.57 \\  \bottomrule
\end{tabular}
\end{table}

\begin{table}[!h]
 \centering
 \caption{Candidates for missions from Mars to Venus.}\label{tab4}
 \vspace*{2ex}
\begin{tabular}{@{}ccccccccc@{}}
\toprule
NEO      & U & Tr. time (days) & Start time  & $\mathrm{r_{1}}$ (km)     & $\mathrm{V_{1}}$ (km $\mathrm{s^{-1}}$) & Finish time & $\mathrm{r_{2}}$ (km) & $\mathrm{V_{2}}$ (km $\mathrm{s^{-1}}$) \\ \midrule
2000 GE2   & 0 & 120 & 2076-Feb-06 & 6861261 & 12.48 & 2076-Jun-05 & 3395856 & 10.35 \\
2013 HT150 & 3 & 100 & 2112-Oct-25 & 9350077 & 16.35 & 2113-Feb-02 & 2650542 & 8.62 \\  \bottomrule
\end{tabular}
\end{table}

\begin{table}[!h]
 \centering
 \caption{Candidates for missions from Venus to Earth.}\label{tab5}
 \vspace*{2ex}
\begin{tabular}{@{}ccccccccc@{}}
\toprule
NEO      & U & Tr. time (days) & Start time  & $\mathrm{r_{1}}$ (km)     & $\mathrm{V_{1}}$ (km $\mathrm{s^{-1}}$) & Finish time & $\mathrm{r_{2}}$ (km) & $\mathrm{V_{2}}$ (km $\mathrm{s^{-1}}$) \\ \midrule
2021 PQ6   & 2 & 40  & 2025-Mar-05 & 9605339 & 10.86 & 2025-Apr-14 & 9121876 & 10.62 \\
2023 KU    & 2 & 30  & 2025-Mar-15 & 5321716 & 20.99 & 2025-Apr-14 & 3735018 & 18.05 \\
2011 EX4   & 1 & 70  & 2029-Dec-19 & 5966989 & 8.45  & 2030-Feb-27 & 5883007 & 6.17  \\
2017 NR6   & 2 & 40  & 2029-Dec-19 & 7451037 & 11.22 & 2030-Jan-28 & 9553408 & 13.21 \\
2006 RJ1   & 0 & 110 & 2031-May-03 & 6885103 & 9.33  & 2031-Aug-21 & 8949121 & 7.54  \\
1936 CA    & 0 & 70  & 2035-Nov-28 & 6307375 & 24.69 & 2036-Feb-06 & 5929391 & 23.78 \\
2006 WB    & 2 & 120 & 2039-Jul-10 & 8201067 & 6.59  & 2039-Nov-07 & 6368956 & 4.09  \\
2018 TP1   & 2 & 40  & 2041-Jan-30 & 4479137 & 10.26 & 2041-Mar-11 & 9619311 & 18.38 \\
2001 SY269 & 1 & 40  & 2041-Feb-09 & 8302257 & 11.21 & 2041-Mar-21 & 4989123 & 16.98 \\
2019 GA    & 1 & 30  & 2041-Mar-01 & 8968917 & 18.44 & 2041-Mar-31 & 9447428 & 17.13 \\
2000 YS134 & 2 & 90  & 2042-Sep-12 & 5957430 & 5.47  & 2042-Dec-11 & 5390159 & 5.39  \\
2001 BB16  & 0 & 160 & 2042-Oct-02 & 3830169 & 4.73  & 2043-Mar-11 & 6918418 & 4.97  \\
2022 QX4   & 2 & 140 & 2045-Apr-09 & 7880047 & 16.66 & 2045-Aug-27 & 3265593 & 8.79  \\
1998 SH36  & 0 & 30  & 2052-May-12 & 9352729 & 17.09 & 2052-Jun-11 & 7914590 & 17.71 \\
2013 SM20  & 0 & 150 & 2055-Mar-18 & 9013393 & 11.30 & 2055-Aug-15 & 7066998 & 5.48  \\
2003 DZ15  & 0 & 80  & 2056-Nov-27 & 6807092 & 11.33 & 2057-Feb-15 & 5425095 & 15.58 \\
2022 MP    & 3 & 140 & 2057-Feb-05 & 6847850 & 9.23  & 2057-Jun-25 & 4433034 & 7.68  \\
2002 LT24  & 0 & 80  & 2057-Apr-06 & 6588250 & 19.56 & 2057-Jun-25 & 9232247 & 8.95  \\
2019 DA    & 1 & 110 & 2061-Oct-22 & 6429161 & 6.02  & 2062-Feb-09 & 9233156 & 5.78  \\
2009 WY7   & 1 & 90  & 2063-Sep-12 & 8647987 & 19.46 & 2063-Dec-11 & 6322086 & 13.98 \\
2021 EL4   & 3 & 110 & 2064-Nov-25 & 7449462 & 9.09  & 2065-Mar-15 & 9494388 & 6.48  \\
2020 UA    & 4 & 150 & 2076-May-16 & 6393315 & 7.25  & 2076-Oct-13 & 6463400 & 5.41  \\
2019 SF6   & 0 & 140 & 2077-May-11 & 8401069 & 16.60 & 2077-Sep-28 & 9003342 & 8.24  \\
2010 FX9   & 0 & 50  & 2081-Jan-30 & 4229118 & 7.02  & 2081-Mar-21 & 6340271 & 11.95 \\
2012 UK171 & 0 & 130 & 2082-Jun-24 & 7303956 & 6.86  & 2082-Nov-01 & 2952527 & 5.35  \\
2012 XS111 & 1 & 30  & 2082-Oct-02 & 6498775 & 17.94 & 2082-Nov-01 & 8430062 & 14.06 \\
2004 SB56  & 0 & 120 & 2087-Mar-30 & 6076922 & 14.43 & 2087-Jul-28 & 8581818 & 11.45 \\
2008 LG2   & 0 & 100 & 2089-Jan-28 & 5617453 & 5.59  & 2089-May-08 & 6603472 & 4.36  \\
2023 UY    & 4 & 110 & 2090-Jul-02 & 9436773 & 8.74  & 2090-Oct-20 & 4591238 & 6.99  \\
2005 YR3   & 1 & 160 & 2093-Jul-26 & 5222651 & 7.84  & 2094-Jan-02 & 8896983 & 5.62  \\
2018 BH3   & 3 & 120 & 2096-Sep-28 & 5418788 & 12.59 & 2097-Jan-26 & 8515351 & 7.97  \\
2017 HG4   & 0 & 100 & 2096-Dec-27 & 1115279 & 4.90  & 2097-Apr-06 & 9164611 & 5.62  \\
2010 RJ43  & 1 & 30  & 2103-Jul-25 & 9016158 & 21.30 & 2103-Aug-24 & 3478165 & 14.80 \\
2003 LN6   & 0 & 130 & 2105-Jan-15 & 6299388 & 5.46  & 2105-May-25 & 3830888 & 3.47  \\
2011 AX22  & 3 & 40  & 2109-Nov-30 & 6945525 & 12.34 & 2110-Jan-09 & 5977168 & 11.40 \\
2019 BE5   & 2 & 30  & 2109-Dec-30 & 7444760 & 24.18 & 2110-Jan-29 & 2419841 & 14.14 \\
2010 KC    & 0 & 30  & 2113-Feb-22 & 6137190 & 21.02 & 2113-Mar-24 & 8780494 & 16.81 \\
1999 GS6   & 0 & 80  & 2116-Jan-08 & 9501083 & 12.73 & 2116-Mar-28 & 7501797 & 14.92  \\ \bottomrule
\end{tabular}
\end{table}

\begin{table}[!h]
 \centering
 \caption{Candidates for missions from Venus to Mars.}\label{tab6}
 \vspace*{2ex}
\begin{tabular}{@{}ccccccccc@{}}
\toprule
NEO      & U & Tr. time (days) & Start time  & $\mathrm{r_{1}}$ (km)     & $\mathrm{V_{1}}$ (km $\mathrm{s^{-1}}$) & Finish time & $\mathrm{r_{2}}$ (km) & $\mathrm{V_{2}}$ (km $\mathrm{s^{-1}}$) \\ \midrule
1998MZ   & 0 & 70  & 2029-Aug-11 & 7109528 & 15.46 & 2029-Oct-20 & 7615254 & 17.00 \\
2019YU3  & 1 & 110 & 2030-Jul-17 & 5814894 & 14.52 & 2030-Nov-04 & 6033618 & 12.51 \\
2010EK43 & 1 & 90  & 2049-Oct-05 & 9041686 & 23.46 & 2050-Jan-03 & 7155966 & 16.25 \\
2018GY3  & 2 & 160 & 2072-Jun-26 & 9361302 & 7.37  & 2072-Dec-03 & 9527195 & 7.56  \\
1999MM   & 0 & 140 & 2090-Sep-10 & 1492449 & 15.36 & 2091-Jan-28 & 5812115 & 14.15 \\
2019UH9  & 1 & 120 & 2105-Mar-26 & 7923342 & 21.92 & 2105-Jul-24 & 7554802 & 16.07 \\  \bottomrule
\end{tabular}
\end{table}

\begin{table}[!h]
 \centering
 \caption{Candidates for orbit refinement (Earth-Venus).}\label{tab7}
 \vspace*{2ex}
\begin{tabular}{@{}ccccccccc@{}}
\toprule
NEO      & U & Tr. time (days) & Start time  & $\mathrm{r_{1}}$ (km)     & $\mathrm{V_{1}}$ (km $\mathrm{s^{-1}}$) & Finish time & $\mathrm{r_{2}}$ (km) & $\mathrm{V_{2}}$ (km $\mathrm{s^{-1}}$) \\ \midrule
2016 VB1   & 6 & 60  & 2037-Nov-17 & 4527976 & 6.37  & 2038-Jan-16 & 6157275 & 12.76 \\
2018 WA1   & 6 & 180 & 2040-Nov-01 & 2958392 & 3.54  & 2041-Apr-30 & 5107153 & 3.59  \\
2015 YA    & 6 & 90  & 2044-May-14 & 3730290 & 8.63  & 2044-Aug-12 & 9432630 & 5.81  \\
2020 VU5   & 6 & 180 & 2045-Oct-26 & 5799337 & 4.78  & 2046-Apr-24 & 5969726 & 5.67  \\
2024 DV    & 6 & 160 & 2047-Feb-18 & 3236436 & 6.32  & 2047-Jul-28 & 9222520 & 10.58 \\
2011 JM5   & 6 & 130 & 2047-Jul-28 & 8191012 & 5.60  & 2047-Dec-05 & 4082413 & 7.85  \\
2016 RL17  & 6 & 110 & 2047-Sep-16 & 7743643 & 9.88  & 2048-Jan-04 & 7855626 & 9.11  \\
2021 RF3   & 6 & 130 & 2050-Sep-10 & 6652138 & 11.00 & 2051-Jan-18 & 2201848 & 14.04 \\
2011 CQ1   & 6 & 140 & 2057-Jan-26 & 4793787 & 4.63  & 2057-Jun-15 & 9324987 & 7.02  \\
2023 DH    & 6 & 130 & 2060-Mar-01 & 4833387 & 4.86  & 2060-Jul-09 & 4293297 & 6.81  \\
2019 UL    & 6 & 180 & 2061-Oct-22 & 1680087 & 3.79  & 2062-Apr-20 & 2201356 & 7.12  \\
2018 BN6   & 6 & 80  & 2063-Jul-24 & 3893967 & 10.80 & 2063-Oct-12 & 7062618 & 6.76  \\
2010 UK    & 6 & 160 & 2064-Oct-26 & 8788090 & 4.88  & 2065-Apr-04 & 6374581 & 6.82  \\
2014 AE29  & 6 & 110 & 2065-Jan-04 & 3138309 & 4.59  & 2065-Apr-24 & 7811177 & 1.76  \\
2023 VA7   & 6 & 110 & 2072-Nov-13 & 3770372 & 6.19  & 2073-Mar-03 & 3232771 & 9.85  \\
2019 AG7   & 6 & 60  & 2073-Jan-12 & 3310678 & 6.50  & 2073-Mar-13 & 8505150 & 10.94 \\
2016 TQ54  & 6 & 70  & 2073-Feb-01 & 9005086 & 7.44  & 2073-Apr-12 & 7229799 & 7.11  \\
2019 UD    & 6 & 110 & 2077-Nov-17 & 7486123 & 5.07  & 2078-Mar-07 & 6482070 & 3.46  \\
2020 WM    & 6 & 30  & 2077-Nov-27 & 4840576 & 11.33 & 2077-Dec-27 & 4718668 & 15.10 \\
2016 CN248 & 6 & 90  & 2078-Feb-15 & 4798146 & 14.24 & 2078-May-16 & 7055120 & 13.77 \\
2023 VN7   & 6 & 90  & 2080-Dec-31 & 9150225 & 4.79  & 2081-Mar-31 & 6819937 & 7.40  \\
2008 CK70  & 6 & 30  & 2081-Feb-19 & 5333821 & 14.67 & 2081-Mar-21 & 2797244 & 14.34 \\
2021 AC2   & 6 & 70  & 2092-Apr-02 & 2587190 & 6.71  & 2092-Jun-11 & 3475005 & 7.82  \\
2003 HT42  & 6 & 100 & 2103-Apr-16 & 4872273 & 5.05  & 2103-Jul-25 & 6642562 & 10.16 \\
2020 DO4   & 6 & 120 & 2103-Jun-05 & 7203840 & 4.39  & 2103-Oct-03 & 2797642 & 4.86 \\  \bottomrule
\end{tabular}
\end{table}

\begin{table}[!h]
 \centering
 \caption{Candidates for orbit refinement (Earth-Mars).}\label{tab8}
 \vspace*{2ex}
\begin{tabular}{@{}ccccccccc@{}}
\toprule
NEO      & U & Tr. time (days) & Start time  & $\mathrm{r_{1}}$ (km)     & $\mathrm{V_{1}}$ (km $\mathrm{s^{-1}}$) & Finish time & $\mathrm{r_{2}}$ (km) & $\mathrm{V_{2}}$ (km $\mathrm{s^{-1}}$) \\ \midrule
2016 EN156 & 6 & 180 & 2035-Mar-23 & 6652303 & 6.53  & 2035-Sep-19 & 8953905 & 6.65  \\
2020 QA6   & 6 & 110 & 2037-Sep-08 & 2522895 & 5.77  & 2037-Dec-27 & 4185081 & 8.16  \\
2018 WG2   & 6 & 160 & 2048-May-03 & 1808018 & 6.36  & 2048-Oct-10 & 5120999 & 4.15  \\
2014 YE1   & 6 & 80  & 2101-Dec-02 & 7688152 & 16.90 & 2102-Feb-20 & 9333846 & 10.01 \\
2020 XY    & 6 & 120 & 2108-Jan-20 & 7605878 & 5.66  & 2108-May-19 & 3510538 & 12.03 \\
2024 BO    & 6 & 100 & 2110-Feb-28 & 5099118 & 5.58  & 2110-Jun-08 & 9478512 & 12.46 \\
2022 EQ    & 6 & 160 & 2114-Mar-09 & 8046760 & 10.02 & 2114-Aug-16 & 3217427 & 11.13 \\  \bottomrule
\end{tabular}
\end{table}

\begin{table}[!h]
 \centering
 \caption{Candidates for orbit refinement (Mars-Earth).}\label{tab9}
 \vspace*{2ex}
\begin{tabular}{@{}ccccccccc@{}}
\toprule
NEO      & U & Tr. time (days) & Start time  & $\mathrm{r_{1}}$ (km)     & $\mathrm{V_{1}}$ (km $\mathrm{s^{-1}}$) & Finish time & $\mathrm{r_{2}}$ (km) & $\mathrm{V_{2}}$ (km $\mathrm{s^{-1}}$) \\ \midrule
2022 WF5   & 7 & 110 & 2033-Jul-31 & 5087386 & 11.34 & 2033-Nov-18 & 7162890 & 10.42 \\
2012 GE    & 7 & 70  & 2035-Jul-11 & 5435221 & 7.40  & 2035-Sep-19 & 3956800 & 11.25 \\
2022 UC34  & 7 & 140 & 2050-Oct-10 & 5174367 & 13.65 & 2051-Feb-27 & 7964233 & 12.33 \\
2023 RK5   & 7 & 110 & 2061-Jan-25 & 8365435 & 8.32  & 2061-May-15 & 3659242 & 9.53  \\
2020 TW2   & 7 & 110 & 2063-Mar-16 & 4794879 & 7.25  & 2063-Jul-04 & 5142872 & 8.38  \\
2016 CL18  & 6 & 170 & 2063-Jul-24 & 6591630 & 8.56  & 2064-Jan-10 & 7594096 & 13.31 \\
2020 UC    & 7 & 90  & 2067-Jul-23 & 7873063 & 6.42  & 2067-Oct-21 & 4208533 & 8.36  \\
2024 FY1   & 7 & 130 & 2067-Oct-31 & 6650386 & 13.58 & 2068-Mar-09 & 8231442 & 9.02  \\
2020 HE1   & 7 & 100 & 2069-Oct-20 & 5385807 & 11.05 & 2070-Jan-28 & 7999562 & 6.64  \\
2024 FD1   & 7 & 130 & 2071-Nov-09 & 3372440 & 10.00 & 2072-Mar-18 & 4404994 & 5.52  \\
2011 EC    & 7 & 120 & 2071-Nov-19 & 8896760 & 10.94 & 2072-Mar-18 & 3934998 & 6.11  \\
2023 FC4   & 7 & 120 & 2074-Jan-17 & 9534181 & 15.13 & 2074-May-17 & 5672467 & 7.61  \\
2014 SB145 & 7 & 90  & 2082-Jul-14 & 8169393 & 6.56  & 2082-Oct-12 & 2364674 & 6.75  \\
2023 ET2   & 7 & 50  & 2082-Aug-03 & 8573837 & 12.63 & 2082-Sep-22 & 5027749 & 12.53 \\
2022 TS2   & 7 & 110 & 2084-Jul-03 & 6100351 & 8.59  & 2084-Oct-21 & 5330754 & 13.43 \\
2019 UT8   & 7 & 80  & 2084-Aug-12 & 5013087 & 9.64  & 2084-Oct-31 & 4015118 & 16.23 \\
2010 WA9   & 7 & 100 & 2084-Sep-11 & 4137542 & 8.16  & 2084-Dec-20 & 7328439 & 6.86  \\
2020 FD3   & 6 & 150 & 2090-Dec-19 & 5534544 & 7.87  & 2091-May-18 & 2361490 & 5.99  \\
2012 XH    & 7 & 120 & 2097-Jul-25 & 8508404 & 8.24  & 2097-Nov-22 & 4125644 & 7.09  \\
2023 TM14  & 7 & 90  & 2097-Aug-14 & 7588675 & 12.40 & 2097-Nov-12 & 8405929 & 7.85  \\
2016 GK2   & 7 & 140 & 2099-Nov-02 & 5079474 & 12.93 & 2100-Mar-22 & 4391983 & 8.29  \\
2020 HS7   & 6 & 80  & 2108-Feb-19 & 8314497 & 13.46 & 2108-May-09 & 9197644 & 13.50 \\
2016 TD    & 7 & 180 & 2108-Apr-09 & 9099941 & 10.64 & 2108-Oct-06 & 4859767 & 13.66 \\
2023 XY2   & 7 & 150 & 2114-Dec-04 & 6339973 & 15.75 & 2115-May-03 & 8206349 & 12.64 \\
2022 GM    & 6 & 140 & 2116-Nov-03 & 9406757 & 12.28 & 2117-Mar-23 & 4352765 & 6.88  \\  \bottomrule
\end{tabular}
\end{table}

\begin{table}[!h]
 \centering
 \caption{Candidates for orbit refinement (Mars-Venus).}\label{tab10}
 \vspace*{2ex}
\begin{tabular}{@{}ccccccccc@{}}
\toprule
NEO      & U & Tr. time (days) & Start time  & $\mathrm{r_{1}}$ (km)     & $\mathrm{V_{1}}$ (km $\mathrm{s^{-1}}$) & Finish time & $\mathrm{r_{2}}$ (km) & $\mathrm{V_{2}}$ (km $\mathrm{s^{-1}}$) \\ \midrule
2020 UE  & 6 & 130 & 2076-Jan-27 & 6505140 & 9.91  & 2076-Jun-05 & 4914913 & 9.77  \\
2023 TE7 & 7 & 140 & 2077-Mar-12 & 6873984 & 14.53 & 2077-Jul-30 & 8698240 & 17.36 \\
2018 TD3 & 7 & 150 & 2079-Aug-29 & 8608976 & 8.25  & 2080-Jan-26 & 8249687 & 12.09 \\
2020 PU4 & 7 & 180 & 2097-Dec-12 & 6765611 & 6.58  & 2098-Jun-10 & 6653926 & 8.19  \\
2020 UQ6 & 7 & 120 & 2110-Feb-08 & 7064027 & 18.41 & 2110-Jun-08 & 7494246 & 19.78 \\  \bottomrule
\end{tabular}
\end{table}

\begin{table}[!h]
 \centering
 \caption{Candidates for orbit refinement (Venus-Earth).}\label{tab11}
 \vspace*{2ex}
\begin{tabular}{@{}ccccccccc@{}}
\toprule
NEO      & U & Tr. time (days) & Start time  & $\mathrm{r_{1}}$ (km)     & $\mathrm{V_{1}}$ (km $\mathrm{s^{-1}}$) & Finish time & $\mathrm{r_{2}}$ (km) & $\mathrm{V_{2}}$ (km $\mathrm{s^{-1}}$) 
\\ \midrule
2014 UN114 & 5 & 140 & 2031-Jun-22 & 4904873 & 4.29  & 2031-Nov-09 & 7964759 & 3.73  \\
2019 EA2   & 5 & 180 & 2045-Sep-16 & 8408276 & 7.13  & 2046-Mar-15 & 5108400 & 4.64  \\
2016 FY2   & 5 & 100 & 2045-Oct-26 & 5486448 & 3.75  & 2046-Feb-03 & 6983267 & 5.50  \\
2010 AF40  & 5 & 70  & 2047-May-19 & 5728754 & 10.21 & 2047-Jul-28 & 6215626 & 13.27 \\
2022 TF1   & 5 & 100 & 2063-Jun-14 & 5305079 & 3.61  & 2063-Sep-22 & 4960038 & 4.69  \\
2023 TM3   & 5 & 110 & 2069-Apr-23 & 4945806 & 17.15 & 2069-Aug-11 & 4846443 & 11.27 \\
2023 VW5   & 5 & 180 & 2073-May-22 & 5862034 & 14.82 & 2073-Nov-18 & 9513081 & 13.34 \\
2018 RQ1   & 5 & 150 & 2084-Apr-04 & 6249327 & 3.14  & 2084-Sep-01 & 3691578 & 2.83  \\
2014 GJ1   & 5 & 110 & 2088-Dec-09 & 5689462 & 6.27  & 2089-Mar-29 & 6881180 & 7.10  \\
2023 RF4   & 5 & 100 & 2090-Jun-02 & 8867285 & 12.09 & 2090-Sep-10 & 1302836 & 11.22 \\
2021 TO8   & 5 & 140 & 2111-May-24 & 5685704 & 13.44 & 2111-Oct-11 & 3230357 & 10.12 \\
2017 FS91  & 5 & 60  & 2113-Jan-23 & 4951136 & 7.10  & 2113-Mar-24 & 1833112 & 7.86 
\\  \bottomrule
\end{tabular}
\end{table}

\begin{table}[!h]
 \centering
 \caption{Candidates for orbit refinement (Venus-Mars).}\label{tab12}
 \vspace*{2ex}
\begin{tabular}{@{}ccccccccc@{}}
\toprule
NEO      & U & Tr. time (days) & Start time  & $\mathrm{r_{1}}$ (km)     & $\mathrm{V_{1}}$ (km $\mathrm{s^{-1}}$) & Finish time & $\mathrm{r_{2}}$ (km) & $\mathrm{V_{2}}$ (km $\mathrm{s^{-1}}$) 
\\ \midrule
2022 MH   & 6 & 180 & 2027-Jul-23 & 7938487 & 11.79 & 2028-Jan-19 & 3584168 & 7.34  \\
2019 PB2  & 7 & 170 & 2091-Aug-26 & 801377  & 8.14  & 2092-Feb-12 & 6022724 & 7.34  \\
2017 XY61 & 7 & 140 & 2089-Sep-25 & 5325640 & 12.90 & 2090-Feb-12 & 1741115 & 11.60
\\  \bottomrule
\end{tabular}
\end{table}

\newpage
\section*{\sc appendix 2}
\begin{table}[!h]
 \centering
 \caption{Candidates for missions Earth-Venus-Mars.}\label{tab13}
 \vspace*{2ex}
\begin{tabular}{@{}ccccccccc@{}}
\toprule
NEO      & U & Tr. time (days) & Start time  & $\mathrm{r_{1}}$ (km)     & $\mathrm{V_{1}}$ (km $\mathrm{s^{-1}}$) & Finish time & $\mathrm{r_{2}}$ (km) & $\mathrm{V_{2}}$ (km $\mathrm{s^{-1}}$) \\ \midrule
2016 DY30 & 7 & 30  & 2079-Jul-20 & 6711331 & 16.89 & 2079-Aug-19 & 7933747 & 14.96 \\
2016 DY30 & 7 & 200 & 2079-Aug-19 & 7933747 & 14.96 & 2080-Mar-06 & 8047451 & 8.47
\\  \bottomrule
\end{tabular}
\end{table}

\begin{table}[!h]
 \centering
 \caption{Candidates for missions Mars-Venus-Earth.}\label{tab14}
 \vspace*{2ex}
\begin{tabular}{@{}ccccccccc@{}}
\toprule
NEO      & U & Tr. time (days) & Start time  & $\mathrm{r_{1}}$ (km)     & $\mathrm{V_{1}}$ (km $\mathrm{s^{-1}}$) & Finish time & $\mathrm{r_{2}}$ (km) & $\mathrm{V_{2}}$ (km $\mathrm{s^{-1}}$) \\ \midrule
2005 TK50 & 9 & 90 & 2080-Nov-01 & 6182474 & 18.38 & 2081-Jan-30 & 6569488 & 11.65 \\
2005 TK50 & 9 & 40 & 2081-Jan-30 & 6569488 & 11.65 & 2081-Mar-11 & 7914667 & 11.22
\\  \bottomrule
\end{tabular}
\end{table}

\begin{table}[!h]
 \centering
 \caption{Candidates for multiple missions from Earth to Venus.}\label{tab15}
 \vspace*{2ex}
\begin{tabular}{@{}ccccccccc@{}}
\toprule
NEO      & U & Tr. time (days) & Start time  & $\mathrm{r_{1}}$ (km)     & $\mathrm{V_{1}}$ (km $\mathrm{s^{-1}}$) & Finish time & $\mathrm{r_{2}}$ (km) & $\mathrm{V_{2}}$ (km $\mathrm{s^{-1}}$) 
\\ \midrule
2014 QX432 & 0 & 30  & 2069-Dec-19 & 12065628 & 16.64 & 2070-Jan-18 & 11938948 & 14.31 \\
2014 QX432 & 0 & 30  & 2077-Dec-17 & 8406818  & 18.08 & 2078-Jan-16 & 9491640  & 15.86 \\
2014 QX432 & 0 & 30  & 2085-Dec-15 & 8047548  & 18.61 & 2086-Jan-14 & 8551610  & 15.80 \\
2014 QX432 & 0 & 30  & 2093-Dec-13 & 8597333  & 18.56 & 2094-Jan-12 & 9590785  & 14.64 \\
2014 QX432 & 0 & 30  & 2101-Dec-12 & 10636625 & 18.37 & 2102-Jan-11 & 14037821 & 13.15 \\
2009 WY7   & 1 & 90  & 2032-Nov-03 & 7482158  & 14.67 & 2033-Feb-01 & 3531904  & 20.79 \\
2009 WY7   & 1 & 70  & 2064-Nov-15 & 8255654  & 12.36 & 2065-Jan-24 & 13092566 & 25.79 \\
2009 WY7   & 1 & 220 & 2074-Dec-03 & 9099317  & 14.67 & 2075-Jul-11 & 6994702  & 21.99 \\
2009 WY7   & 1 & 20  & 2085-Dec-05 & 12545349 & 15.79 & 2085-Dec-25 & 12975557 & 19.05 \\
2009 WY7   & 1 & 170 & 2087-Nov-05 & 12633364 & 15.80 & 2088-Apr-23 & 13106231 & 18.85 \\
2019 SF6   & 0 & 50  & 2045-Nov-05 & 8356719  & 8.45  & 2045-Dec-25 & 9111880  & 14.46 \\
2019 SF6   & 0 & 130 & 2047-Nov-05 & 8052266  & 8.61  & 2048-Mar-14 & 13212386 & 17.67 \\
2019 SF6   & 0 & 50  & 2053-Nov-13 & 7419557  & 8.42  & 2054-Jan-02 & 8353232  & 14.16 \\
2019 SF6   & 0 & 130 & 2055-Nov-13 & 8415482  & 8.27  & 2056-Mar-22 & 15239381 & 18.51 \\
2018 RY1   & 0 & 210 & 2050-Aug-11 & 5354993  & 7.41  & 2051-Mar-09 & 4191982  & 13.09 \\
2018 RY1   & 0 & 100 & 2053-Sep-04 & 13855613 & 6.55  & 2053-Dec-13 & 4024853  & 12.82 \\
2018 RY1   & 0 & 210 & 2082-Aug-13 & 8964879  & 6.65  & 2083-Mar-11 & 8823827  & 14.48 \\
2018 RY1   & 0 & 100 & 2085-Sep-06 & 14211426 & 6.51  & 2085-Dec-15 & 5283290  & 10.39 \\
2023 JK3   & 5 & 200 & 2036-Apr-16 & 2300432  & 4.99  & 2036-Nov-02 & 8933252  & 12.01 \\
2023 JK3   & 5 & 150 & 2044-May-24 & 2262544  & 4.67  & 2044-Oct-21 & 6085544  & 7.41  \\
2023 JK3   & 5 & 170 & 2052-May-12 & 2372499  & 4.49  & 2052-Oct-29 & 4350493  & 9.50  \\
2023 JK3   & 5 & 100 & 2079-May-01 & 3390197  & 4.47  & 2079-Aug-09 & 3026153  & 8.66  \\
2011 EP51  & 0 & 150 & 2057-Mar-07 & 11146314 & 6.68  & 2057-Aug-04 & 9311846  & 8.53  \\
2011 EP51  & 0 & 100 & 2060-Feb-20 & 14060754 & 6.52  & 2060-May-30 & 9170173  & 8.69  \\
2011 EP51  & 0 & 60  & 2100-Mar-22 & 6185111  & 7.11  & 2100-May-21 & 5606121  & 11.66 \\
2005 VL1   & 4 & 70  & 2033-Feb-01 & 3338198  & 5.94  & 2033-Apr-12 & 5108226  & 6.74  \\
2005 VL1   & 4 & 70  & 2049-Feb-07 & 2455822  & 6.02  & 2049-Apr-18 & 3272321  & 5.68  \\
2005 VL1   & 4 & 180 & 2091-Nov-04 & 6641524  & 5.11  & 2092-May-02 & 8765801  & 8.30 \\  \bottomrule
\end{tabular}
\end{table}

\begin{table}[!h]
 \centering
 \caption{Candidates for multiple missions from Venus to Earth.}\label{tab16}
 \vspace*{2ex}
\begin{tabular}{@{}ccccccccc@{}}
\toprule
NEO      & U & Tr. time (days) & Start time  & $\mathrm{r_{1}}$ (km)     & $\mathrm{V_{1}}$ (km $\mathrm{s^{-1}}$) & Finish time & $\mathrm{r_{2}}$ (km) & $\mathrm{V_{2}}$ (km $\mathrm{s^{-1}}$) 
\\ \midrule
2019 SF6  & 0 & 70  & 2071-Aug-01 & 11841784 & 11.77 & 2071-Oct-10 & 13681512 & 8.10  \\
2019 SF6  & 0 & 140 & 2077-May-11 & 9003342  & 16.60 & 2077-Sep-28 & 8401069  & 8.24  \\
2019 SF6  & 0 & 60  & 2079-Jul-30 & 8317517  & 16.24 & 2079-Sep-28 & 10275580 & 8.35  \\
2019 SF6  & 0 & 140 & 2085-May-09 & 7545164  & 14.90 & 2085-Sep-26 & 8718373  & 8.45  \\
2019 SF6  & 0 & 60  & 2087-Jul-28 & 7872182  & 17.74 & 2087-Sep-26 & 13317101 & 8.39  \\
2019 SF6  & 0 & 140 & 2093-May-17 & 9902219  & 15.44 & 2093-Oct-04 & 8544228  & 8.31  \\
2019 SF6  & 0 & 70  & 2095-Jul-26 & 10572438 & 16.09 & 2095-Oct-04 & 11102066 & 8.16  \\
2019 SF6  & 0 & 160 & 2101-May-06 & 12397691 & 18.37 & 2101-Oct-13 & 13223363 & 8.05  \\
2019 SF6  & 0 & 90  & 2103-Jul-25 & 12907143 & 12.41 & 2103-Oct-23 & 12833372 & 8.04  \\
2009 WY7  & 1 & 70  & 2031-Sep-20 & 8486240  & 23.79 & 2031-Nov-29 & 11324651 & 12.04 \\
2009 WY7  & 1 & 90  & 2063-Sep-12 & 6322086  & 19.46 & 2063-Dec-11 & 8647987  & 13.98 \\
2009 WY7  & 1 & 130 & 2075-Jul-11 & 8523302  & 21.99 & 2075-Nov-18 & 6994702  & 12.12 \\
2009 WY7  & 1 & 30  & 2098-Oct-08 & 9675764  & 21.79 & 2098-Nov-07 & 6036873  & 14.69 \\
2023 TM3  & 5 & 190 & 2034-Mar-28 & 8886083  & 14.90 & 2034-Oct-02 & 9973498  & 9.66  \\
2023 TM3  & 5 & 200 & 2058-Mar-22 & 2219710  & 17.62 & 2058-Oct-08 & 3176879  & 11.15 \\
2023 TM3  & 5 & 110 & 2069-Apr-23 & 4846443  & 17.15 & 2069-Aug-11 & 4945806  & 11.27 \\
2023 TM3  & 5 & 130 & 2116-Jun-06 & 3194864  & 13.54 & 2116-Oct-14 & 14941647 & 11.17 \\
2011 EX4  & 1 & 70  & 2029-Dec-19 & 5966989  & 15.70 & 2030-Feb-27 & 5883007  & 23.23 \\
2011 EX4  & 1 & 100 & 2093-Dec-13 & 13820620 & 9.02  & 2094-Mar-23 & 5674572  & 5.62  \\
2011 EX4  & 1 & 100 & 2104-Nov-06 & 10138626 & 6.72  & 2105-Feb-14 & 15198492 & 9.10 \\  \bottomrule
\end{tabular}
\end{table}

\begin{table}[!h]
 \centering
 \caption{Candidates for multiple missions from Venus to Mars.}\label{tab17}
 \vspace*{2ex}
\begin{tabular}{@{}ccccccccc@{}}
\toprule
NEO      & U & Tr. time (days) & Start time  & $\mathrm{r_{1}}$ (km)     & $\mathrm{V_{1}}$ (km $\mathrm{s^{-1}}$) & Finish time & $\mathrm{r_{2}}$ (km) & $\mathrm{V_{2}}$ (km $\mathrm{s^{-1}}$) 
\\ \midrule
2019 YU3 & 1 & 110 & 2030-Jul-17 & 5814894  & 14.51 & 2030-Nov-04 & 6033618  & 12.50 \\
2019 YU3 & 1 & 110 & 2062-Oct-27 & 7582414  & 13.03 & 2062-Jul-09 & 10463241 & 12.33 \\
2019 YU3 & 1 & 130 & 2094-Oct-09 & 12221689 & 12.43 & 2094-Jun-01 & 7112363  & 12.12 
 \\  \bottomrule
\end{tabular}
\end{table}

\end{document}